
\magnification=1200
\pretolerance=10000
\baselineskip=15pt
\headline{\hfill IAG/USP Report No 35/95}
\centerline {\bf DETECTABILITY OF GRAVITATIONAL WAVE BURSTS FROM A}
\centerline {\bf CLASS OF NEUTRON STAR STARQUAKE GRB MODELS}
\vskip 1 true cm
\centerline {J.E.Horvath}
\centerline {\it Instituto Astron\^omico e Geof\'\i sico}
\centerline {\it Universidade de S\~ao Paulo}
\centerline {\it Av. M. St\'efano 4200 - Agua Funda}
\centerline {\it (04301-904) S\~ao Paulo - SP - Brasil}
\vskip 3 true cm
\noindent
{\bf Abstract}
A large class of gamma-ray burst (GRB) models (overwhelming until recently)
involve the release of energy in a neutron star quake. Even though the extreme
isotropy of the GRB sky established by the BATSE experiment has now shifted
the interest to cosmological models, the former starquake scenarios are still
attractive and may naturally produce a gravitational wave burst which carries
most of the released energy. We discuss the prospects for detection of these
high-frequency bursts by the forthcoming interferometers and spheroidal
antennas, emphasizing the most recent results on the distribution and nature
of the GRB sources. We find that, even if the overall picture is correct,
the positive detection of GRB-associated gravitational wave bursts is
unlikely and therefore these events will not be a prime target for the
detectors.

\vskip 3 true cm
\centerline{To appear in {\it Int.J.Mod.Phys.D}}
\vfill\eject

\noindent
{\bf 1. Introduction}
\vskip 0.5 true cm
More than 20 years after their discovery$^{1}$, $\gamma$-ray bursts (hereafter
GRB) continue to puzzle the astrophysics community. In spite of the
availability of an expanded data set and an intense activity on theoretical
modeling$^{2}$, there is no general consensus about the source/sources of
these events. A main ingredient for this confusing situation is the lack
of firmly established counterparts at any other wavelenght, which leaves
the distance scale totally undetermined for an observed highly isotropic
(but inhomogeneous) distribution$^{3}$. In fact, the distance scale has been
postulated to be as short as $10^{4}$ AU (Oort cloud scale) and as long as
several Gpc (cosmological scale). At first glance it seems that the evidence
points more strongly towards a cosmological origin of the bursts since in
that case isotropy is a naturally expected feature. However, prior to
the launch of BATSE experiment onboard the Compton Observatory$^{3}$, several
lines of reasoning leaded to the widespread belief that galactic neutron stars
(hereafter NS) were the sources of the bursts. In fact, some kind of violent
disturbance in a NS continues to be an attractive model for the events
although, generally speaking, the typical distance to a burster had to be
increased in order to satisfy the isotropy constraints, giving rise to the
so-called "extended halo/corona" distributions.

Several lines of attack are being pursued to solve this modern version of
galactic vs. extragalactic controversy. They include searches for
"cosmologically stretched" bursts $^{4}$, searches for
repeating sources$^{5,6}$
and a multi-wavelenght monitoring of the error boxes in real time$^{7}$.
The purpose of this work is to discuss the prospects for detecting a few
bursts/yr at the forthcoming LIGO-type interferometers$^{8,9}$ and
spheroidal resonant antennas$^{10}$. Even though it is not unlikely that
important new evidence to solve the mistery becomes available prior to the
implementation of these facilities in more "conventional" wavelenghts
rather than gravitational waves (GW), the significance of a positive detection
from them would be such an unique opportunity to learn about NS structure
and GW themselves that its importance can not be overstated.

\vfill\eject

{\bf 2. NS models of GRBs}
\vskip 0.5 true cm
In its most popular and widespread version, a NS model of GRB needs the sudden
release of a substantial amount of stored energy, generally associated with
the cracking of a strained solid crust. The propagation of waves then shake the
frozen field lines that then radiate energetic photons. Possible scenarios
have been addressed in Refs.11 and 12, see also Ref.13 for a recent review.
As shown in Ref.14, even a detailed treatment of the involved physics
does not dissipate several uncertainties inherent to this model. Calculations
indicate that up to $\sim \, 10^{44} \, erg$ may be stored as elastic energy
in the crustal lattice. The energy released per quake $\Delta E$ producing
a GRB can be estimated as$^{14}$

$$ \Delta E \, = \, 10^{38} \, \eta^{-1}
{\biggl( {F \over {10^{-6} \, erg \, cm^{-2}}} \biggr)}
{\biggl( {r \over {1 \, kpc}} \biggr)}^{2} \; erg \, = \,
{\Delta E_{\gamma} \over {\eta}} \eqno(1) $$

where $\eta$ is the (unknown) efficiency of the conversion of the energy
into $\gamma$ rays which will be useful to parametrize our discussion,
and $F$ is a typical burst fluence for a source located
at a distance $r$.

After the quake shear and compression waves will propagate through the
star. As discussed in Ref.14, a force-free configuration like the
quaking NS will partition the energy as the inverse of the sixth-power of
the ratio of the shear and longitudinal sound speeds. This is analogous
to the Earth case$^{15}$ and results from the fact that quadrupoles are the
lowest modes for the force-free problem. The typical frequency at which
quadupolar oscillations will produce GW that damp out this motion
depends on the exact composition of the matter at densities above the
nuclear saturation one. Numerical computation of those frequencies have
been performed by Thorne and coworkers$^{16}$ and more recently by
Lindblom and Detweiler$^{17}$ for a complete set of equations of state.
We have chosen below to scale the estimates to the values of a
$1.43 \, M_{\odot}$ Bethe-Johnson I NS model (see Ref.17 and references
therein), having a
quadrupole frequency $f_{c} \, \simeq \, 2 \, kHz$. The frequencies
arising from other choices of the equation of state differ somewhat from
this value (for the given mass) only if matter is substantially stiffer
or softer than the former, and the corresponding GW strenghts can be
easily found if necessary. We have made no attempt to compute the actual
GW waveform to be expected since this is likely to be plagued by the same
uncertainties affecting the generation of GRB (see Ref.14 for a through
discussion of these issues). For our purposes it is sufficient to
adopt the expression of the characteristic amplitude of the waves $h_{c}$
given in Ref.18.

$$ h_{c} \, = \, 2.7 \, \times \, 10^{-17} \,
{\biggl( {\Delta E_{GW} \over {M_{\odot} \, c^{2}}} \biggr)}^{1/2} \,
{\biggl( {1 \, kHz \over {f_{c}}} \biggr)}^{1/2} \,
{\biggl( {10 \, kpc \over {r}} \biggr)} \eqno(2) $$

which is given in terms of the characteristic frequency $f_{c}$ and the
energy put in gravitational waves $\Delta E_{GW}$ with the corresponding
distance factor $\propto \, 1/r$ arising from the quadrupolar character
of the emission.

 We turn now to a brief characterization of the detectors, addressing GW
interferometers first. As discussed in Refs. 19 and 20 the important
quantity that should be calculated for the detection (besides the value of
$h_{c}$) is the signal-to-noise ratio. For a LIGO-type broad-band
interferometer the latter reads

$$ {S \over {N}} \, = \, {h_{c} \over {h_{n}(f_{c})}} \eqno(3) $$

for an assumed optimal filtering. Here $h_{n}(f_{c})$ is the characteristic
detector noise amplitude evaluated at $f_{c}$. As shown in Ref.18, the
expected sensitivities of the advanced generation to bursts is limited by
photon shot-noise in the high-frequency region where
$h_{n}(f_{c}) \, \propto \, f_{c}$. Inserting numbers and using eq.(3) we
obtain

$$ {S \over {N}} \, = \, 10^{5} \,
{\biggl( {\Delta E_{GW} \over {M_{\odot} \, c^{2}}} \biggr)}^{1/2} \,
{\biggl( { 1 \, kHz \over {f_{c}}} \biggr)}^{3/2} \,
{\biggl( { 10 \, kpc \over {r}} \biggr)} \eqno(4) $$

or better, depending on several possible technological improvements under
study that may increase the overall coefficient.

Up to this point we have made no use of the quaking NS hypothesis but merely
restated known results with an appropiate scaling. We specialize now for the
case of NS quakes by identifying
$\Delta E_{GW} \, \leq \, \Delta E \, = \, \Delta E_{\gamma}/\eta$
and write eqs.(2) and (4) as

$$ h_{c} \, = \, 2.7 \, \times 10^{-17} \, \eta^{-1/2} \,
{\biggl( {\Delta E_{\gamma} \over {M_{\odot} \, c^{2}}} \biggr)}^{1/2} \,
{\biggl( { 1 \, kHz \over {f_{c}}} \biggr)}^{1/2} \,
{\biggl( {10 \, kpc \over {r}} \biggr)} \eqno(5) $$

$$ {S \over {N}} \, = \, 10^{5} \, \eta^{-1/2} \,
{\biggl( {\Delta E_{\gamma} \over {M_{\odot} \, c^{2}}} \biggr)}^{1/2} \,
{\biggl( { 1 \, kHz \over {f_{c}}} \biggr)}^{3/2} \,
{\biggl( {10 \, kpc \over {r}} \biggr)}. \eqno(6) $$

Since $\eta$ is a small number (of the order of a few percent at most),
most of the energy comes out in GW for an oscillating star$^{11}$. In
what follows, we shall consider the sensitivity of an advanced LIGO
detector$^{19}$ and consider $S/N \, > \, 2$ as a reasonable criterion for
detection$^{20}$. Optimal filtering and orientation of the interferometer(s)
are also assumed as done in previous works. Before addressing the
astrophysical setting of the sources we should remark that the GRB
detection offer in principle important advantages when compared to other
possible GW bursts. Since the gammas
would act as an electromagnetic counterpart
of the GW signal, they could be identified below the sensitivity threshold
calculated for a random distribution of the bursts$^{19}$ (i.e. those not
associated with a GRB error box). This results in an additional factor
$\sim \, 3$ already taken into account in the above expressions for
$h_{c}$ and $S/N$. The position of the GRB would be a powerful check for
the independently determined position of the GW by two or more detectors,
with the corresponding gain of physical information about the sorces and
the waveform itself.

An alternative and very promising complementary technique to investigate
the high-frequency range is the recent proposal of several spheroidal
resonant antennas$^{10,21}$. The goals of this "$4^{th}$" generation of
detectors cooled to $T \, \leq \, 0.1 \, K$ are to reach high sensitivities
of $h_{c} \, \sim \, 10^{-21}$ or better in a short construction
timescale. In particular, it is expected that a truncated icosahedron design
can be not only $\sim \, 50$ times more sensitive than a bar antenna with
the same noise temperature, but also that the direction and polarization
of the wave can be measured at once on each of these systems. According
to the calculations of Ref.22 an $1.3 \, m$ diameter buckyball of $Al$
having a central frequency of $2 \, kHz$ and a bandwidth
$\Delta f \, \sim \, 100 \, Hz$ would have a strain noise spectrum one
order-of-magnitude lower than an optimally oriented first-generation
LIGO-type interferometer and comparable to the advanced generation
estimates
(other materials and geometries may be even more useful and are under study).
This sensitivity will result in a similar $S/N$ ratio than the
eq.(6) one, but given the potential operability in $\sim \, 3 \, yr$ because
of fewer foreseen technological problems$^{22,23}$ they may become available
faster than the long-shot advanced LIGOs. In the remaining of this work
we shall assume that either technology will be finally able to observe
the possible GRB-associated emission.

\vskip 0.5 true cm
\noindent
{\bf 3. Detectability of the GW from GRB burst sources }
\vskip 0.5 true cm
As stated in the Introduction, the confirmed isotropy of the GRB sky diminished
the confidence the researchers had prior to BATSE launch about the correctness
of the NS picture. However, for a variety of reasons NS should still be
considered as likely sources. In fact, many variants of the latter model have
been constructed$^{24}$ and it has been claimed that the data is indeed
consistent with a repeating population of galactic NS$^{25}$. Previous analysis
of the possible GW events produced by vibrating NS can be
found in Refs.26 and 27,
although these works did not address recent GRBs models but rather
concentrated on the general features of the quakes. We
shall discuss the GW detection of several subclasses of GRBs
following the most recent
advances in the understanding of the latter.

\bigskip
\noindent
a) $"Classical" \, GRB$ : The availability of BATSE data on the dipole and
quadrupole moments of the classical GRB distribution has challenged the
view of model builders. Both quantities  {\bf D} $= \, < cos \, \theta >$ and
{\bf Q} $= \, < sin^{2} b > \, - \, 1/3$ are amazingly close to zero and
since the brightness distribution requires a decrease of the number of
sources with distance$^{3}$, it is difficult to model them as a known
component of the galactic disk which requires the typical distance to the
sources $d$ to be less than the scale height of the disk $Z_{D}$ to
account for the data. However, there have been claims that this is the case
and that GRB are associated with the galactic arms at $\sim \, 1 \, kpc$
distance scale$^{25}$. In such a case, there would be no need of
extended halo/extragalactic sources to explain the events. Given the
typical fluence $F \, = \, 10^{-6} \, erg \, cm^{-2}$ of a GRB we get,
according to eq.(1)
$\Delta E_{GW} \, \simeq \, 10^{38} \, \eta^{-1} \, erg$ and
therefore $S/N \, = \, 3 \, \times \, 10^{-3} \, \eta^{-1/2}$. Thus, a
burst would be detectable (i.e. $S/N \, > \, 2$) if
$\eta \, \leq \, 3 \, \times \, 10^{-6}$.
In other words the actual efficiency of
the conversion to gammas must be very low for the associated GW to be detected,
so low that in the latter case each event must release at least
$\Delta E \, \geq \, 3 \, \times \, 10^{43} \, erg$
or about $10 \, \%$ of
the total elastic energy
stored in the crust.
This is a very severe requirement
since, if a local galactic
population alone is invoked, the observed GRB rate
of $1/day$ calls for at least $10^{5}$
bursts/NS over a Hubble lifetime of
the object. It is apparent that if this is the case, the quantity $\Delta E$
must be much lower (higher $\eta$) and
therefore the GRB events will not be seen
by the forthcoming detectors.

There is, however, another popular
modeling of the sources involving
NS in which an extended halo/corona
is the main responsible for the
isotropy without excluding a disk
NS contribution. The models of Higdon and
Lingenfelter$^{28}$ and Smith and Lamb$^{29}$
are examples of dual populations.
Li et al.$^{30}$ propose that high-velocity
NS populate the halo and produce
the events. Hakkila et al.$^{31}$ and
Smith$^{32}$ have shown that a significant
fraction of the sources (up to $30 \, \%$)
can be in the disk, so that in
these cases the energy problem discussed
above may be avoided since the disk
NS would not be required to reproduce
the whole distribution. But even
if this is the case it is not
automatically guaranteed that the GRB from
the disk can produce GW signals at an intersting rate
$\sim \, 10 $ events/yr. Let us assume
that $\eta$ is much higher, say
$\geq \, 10^{-3}$ as seems reasonable. If so
$\Delta E \, \sim \, 10^{41} \, erg$ and
with the same criterion given above
for a positive detection we obtain that the
distance to the source can not
be larger than $r_{m} \, \simeq \, 40 -50 \, pc$.
The closest NS out of the
$10^{9}$ present in the galaxy is likely to
be $\sim \; 10 \, pc$ away$^{33}$
and therefore there are at least 100 potential sources
in a sphere of radius $r_{m}$.
Thus, since we are sampling
$few \, \times \, 10^{6}$ out to $\sim \, kpc$ scale and
in these composite models the latter
can produce $\sim \, 10 \, \%$ of the
annual events, the probability of observing
a burst closer than $50 \, pc$
(so that it can be also detected in GW)
is $P \, \sim \, 10^{-3}$. Even
if the Quashnock-Lamb results hold and
the whole disk NS population$^{34}$ adds up
to produce the events, $P$ increases to a
meager $2 \, \%$. Needless to say,
these are not very encouraging numbers.

The best prospects for GW detection arise if the very intense events like
GRB 910601 having $F \, = \, 5 \, \times \, 10^{-5}$
are simply the closest to the Earth from an extended
halo/corona distribution.
{}From eq.(1) we see that its distance should be about $0.15 \, R$,
( where $R$ is the typical distance to a souce $\sim \, 10-20 \, kpc$ in
these models).
In such a case the associated GW may be detectable if $\eta \, < \, 10^{-3}$
which is small but perhaps not unreasonable.

\bigskip
\noindent
b) $"Soft" \, \, Gamma \, \,Repeaters$

According to most researchers, the so-called soft-gamma repeaters
(hereafter SGR) represent a subclass of $\gamma$ transients differing
from their "classical" cousins because they present

\noindent
i) Stochastic recurrence patterns and short repetition times

\noindent
ii) Average duration peaked at $\sim \, 1 \, s$.

\noindent
iii) Constant spectral shapes with maximum output at $E \, \sim \, 30 \, keV$.

\noindent
iv) Lack of substantial spectral evolution.

\noindent
v) Rapid rise $and$ decay timescales, unresolved in most cases.

Only three repeating sources heve been identified as such, notably
SGR 0526-66 coincident with the position of the celebrated 1979 March 5
superburst$^{35}$. We discuss this association first, having in mind the
widespread (but not necessarily correct) picture that superbursts are
a manifestation of an internal phenomenon (e.g. phase transitions$^{36}$)
that triggers and active period of SGR of the source.

\bigskip
\noindent
$\ast$ {\it 1979  March  5  event  (SGR 0526-66)}

This is the most celebrated GRB event recorded so far, although it is not
considered itself as a part of the SGR class. Among its unique features,
the exceedingly large fluence allowed a detection by 12 instruments, making
possible a quite precise localization. Its association with the LMC supernova
remnant N49$^{35}$ was subsequently debated until recently (see below), but
there is now firm evidence for a LMC origin at $D \, = \, 50 \, kpc$
strenghtened by
the identification with supernova remnants of the remaining two sources.
The position is also consistent with the source $SGR \, 0526-66$, very
suggestive of a scenario in which $GRB \, 050379$ triggered an "active"
period of the former, identified with a young NS.
Assuming an isotropic emission of $GRB \, 050379$, the detected fluence implies
$\Delta E_{\gamma} \, \geq \, 10^{44.8} \, erg$. Our estimation for GW
is then $S/N \, = \, 0.132 \; \eta^{-1/2}$; suggesting detectability of
the burst if $\eta \, \leq \, 5  \, \times \, 10^{-3}$. Even though there is a
considerable uncertainty on the nature of the event and the precise form of
the $\gamma$ flash generation, the constraint on $\eta$ seems not too strong.
Furthermore, Ramaty et al.$^{36}$ have demonstrated that the temporal history
of the event is compatible with GW damping of a vibrating NS but the
elastic crust energy considerations do not apply since the free energy source
is likely to be associated with core phase transitions. However, it
should be remembered that the event remains unique in more than 20 years of
GRB observations and thus the frequency at which the galaxy and local
extragalactic neighbours produce potentially detectable events disfavors
them as promising targets.

\bigskip
\noindent
$\ast$  {\it SGR 1806-20, SGR 1900+14}  and {\it SGR 1806-20}

These three identified sources of SGR have attracted lots of attention since
the identification with supernovae remnants$^{37,38}$ and the confirmation
by the GINGA satellite team$^{39}$. It now appears that the distance scale is
rather well established ($D \, \sim \, 15 \, kpc$ for $SGR \, 1806-20$ in
SNR G10.0-0.3 .
and $D \, \sim \, 50 \, kpc$ for $SGR \, 0526-66$ in $N49$) and the hypothesis
of a quaking NS leftover from the explosions seems reasonable$^{40}$, even
though alternative explanations invoking abnormally high magnetic fields have
been put forward$^{41}$ so that vibration of the star is not compellingly
involved.
Adopting $15 \, kpc$ as the actual distance we get
$\Delta E \, \simeq \, 10^{41} \, \eta^{-1} \, erg$
for the strongest bursts of $SGR \, 1806-20$ having
$F \, \sim 3 \, \times \, 10^{-5} \, erg \, cm^{-2}$. Our estimate is then
$S/N \, = \, 6 \, \times \, 10^{-3} \, \eta^{-1/2}$ and the GW bursts
possibly associated with the sources will be detectable if
$\eta \, \leq \, 8 \, \times \, 10^{-6}$. The total energy that must
be radiated per event is well above the limiting value that can be stored
in the crust (and this figure must yet be multiplied by the total number of
events for a given source). Even without considering the emission of GW
(i.e. setting $\eta \, \simeq \, 1$) the total energy radiated in GRB happens
to be greater than the elastic crust value and the latter has to be
replenished to match the energetic requirements of the observed active periods.
These numbers suggest that although NS are strong candidates for SGR origin,
it is unlikely that quakes can provide an explanation for them, and thus
GW bursts need not to arise at any intensity level after all.

\vskip 0.5 true cm
\noindent
{\bf 4. Conclusions}
\vskip 0.5 true cm
We have discussed in this work the detectability of GW bursts possibly
associated with a definite model of GRB generation, namely the quakes of NS.
This model has been paradigmatic until recent recent results announced by
the BATSE team established clearly that, even though a fraction of them
may originate in this fashion, it is unlikely that the model can provide
a full explanation for the data by itself. This fact shifted the interest
to cosmological alternatives, but the debate is not likely to end soon.
We have tried to express our estimates as closely related as possible to
these new GRB data. Our conclusion is that "classical" GRB, assumed to
be produced by NS quakes and allowed to be up to $30 \, \%$ of the whole
distribution may be seen in future GW detectors if one of the burst sources
lies within $50 \, pc$, but the chance probability of such an event is
$\sim \, 10^{-3}$ ($0.02$ in the extreme case that all them are galactic
as advocated by Quashnock and Lamb$^{25}$). There may be a reasonable
prospect for GW detection if most of the classical bursts arise form
extended halo/corona distributions and the very intense ones are the
closest $\sim$ few $kpc$ away. Events like the famous 1979 March 5
would be observable by LIGO-type interferometers and buckyball arrays, but
they seem to be too rare to produce a significative rate. Finally, we found
that SGR, now known to be associated with young NS are not likely to be
detected unless our knowledge of the elastic properties of dense matter
is grossly wrong and those compact objects are more exotic than we
think$^{42}$.
A similar analysis of GRB-GW coincidence for the popular NS-NS inspiraling
cosmological sources has been made by Nicholson and Schutz$^{43}$.
These type of works represent the first attempts towards a "multi-wavelenght"
study of high-energy phenomena that will be greatly stimulated by GW
detectors operation in the near future.

\vskip 0.5 true cm
\noindent
{\bf Acknowledgements}

We acknowledge Dr. O.D. Aguiar for his interest in this problem and
valuable information and advice. The financial
support of the Conselho Nacional de
Desenvolvimento Cient\'\i fico e Tecnol\'ogico (CNPq) , Brazil is also
acknowledged.
\vfill\eject
\noindent
{\bf References}
\bigskip
\noindent
1.  R.W.Klebesadel, I.B.Strong and R.A.Olson, {\it Astrophys.J.Lett.}
{\bf 182}, L85 (1973).

\bigskip
\noindent
2. R.Nemiroff, in {\it Proceedings of the $2^{nd}$ Huntsville Workshop on
Gamma-Ray Bursts}, AIP Conf. Proc. {\bf 307},
G.Fishman,J.Branierd and K.Hurley (Eds.), 730 (1994).

\bigskip
\noindent
3. C.Meegan et al., {\it Nature} {\bf 355}, 143 (1992).

\bigskip
\noindent
4. S.P.Davies et al., in Ref.2, 182.

\bigskip
\noindent
5. J.M.Quashnock and D.Q.Lamb, {\it Mon.Not.R.A.S.} {\bf 265}, L69 (1993).

\bigskip
\noindent
6. V.C.Wang and R.E.Lingenfelter, in Ref.2, 160.

\bigskip
\noindent
7. S.D.Barthelmy et al., in Ref.2, 643.

\bigskip
\noindent
8. A.Giazzoto, {\it Phys. Rep.} {\bf 182}, 367 (1989).

\bigskip
\noindent
9. R.Vogt, talk given at
the $12^{th}$ {\it International Conference on General Relativity and
Gravitation  (GR12)}.

\bigskip
\noindent
10. See, for example,
O.D.Aguiar et al., in {\it Proc. of the $13^{th}$ International Conference on
General Relativity and Gravitation (GR13)}, W.Lamperti and O.E.Ortiz (Eds.),
455 (1993).

\bigskip
\noindent
11. A.I.Tsygan, {\it Astron.Astrophys.} {\bf 44}, 21 (1975) ;
A.C.Fabian, V.Icke and J.E.Pringle, {\it Astrophys. Space Sci.} {\bf 42},
77 (1976).

\bigskip
\noindent
12. F.Pacini and M.Ruderman, {\it Nature} {\bf 251}, 399 (1974) ; M.Ruderman,
{\it Astrophys. J.} {\bf 382}, 587 (1991).

\bigskip
\noindent
13. E.P.Liang, in {\it Proceedings of the GRO Science Workshop}, N.Johnson
(Ed.), 397 (1989).

\bigskip
\noindent
14. O.Blaes, R.D.Blandford and P.Goldreich, {\it Astrophys.J.} {\bf 343}, 839
(1989).

\bigskip
\noindent
15. T.Kasawara, {\it Earthquake Mechanics}, (J.Wiley \& Sons, NY 1981).

\bigskip
\noindent
16. K.S.Thorne and A.Campolattaro, {\it Astrophys.J.} {\bf 149}, 591 (1967) ;
R.Price and K.S.Thorne, {\it Astrophys.J.} {\bf 155}, 163 (1969) ;
K.S.Thorne and A.Campolattaro, {\it Astrophys.J.} {\bf 159}, 847 (1970).

\bigskip
\noindent
17. L.Lindblom and S.Detweiler, {\it Astrophys.J.Supp.} {\bf 53}, 73 (1983).

\bigskip
\noindent
18. K.S.Thorne, in {\it 300 Years of Gravitation}, S.Hawking and W.Israel
(Eds.), 330 (Cambridge University Press, London 1987).

\bigskip
\noindent
19. A.Abramovici {\it et al.}, {\it Science} {\bf 256}, 325 (1992).

\bigskip
\noindent
20. D.Kennefick, {\it Phys. Rev. D} {\bf 50}, 3587 (1994).

\bigskip
\noindent
21. W.O.Hamilton and W.W.Johnson, {\it Proposal for a Gravitational Wave
Observatory}, proposal to the National Science Foundation (1990) ; O.D.
Aguiar {\it et al.}, communication to the
{\it Proceedings of the Cryogenic Gravitational Wave
Antenaae Progress Workshop}, Legnaro, Italy (1993).

\bigskip
\noindent
22. W.W.Johnson and S.M.Merkowitz, {\it Phys.Rev.Lett.} {\bf 70}, 2367 (1993).

\bigskip
\noindent
23. O.D.Aguiar, private communication.

\bigskip
\noindent
24. D.H.Hartmann, in Ref.2, 562.

\bigskip
\noindent
25. J.M.Quashnock and D.Q.Lamb, {\it Mon.Not.R.A.S.} {\bf 265}, L45 (1993).

\bigskip
\noindent
26. K.S.Thorne, in {\it Theoretical Principles in Astrophysics and Relativity}
N.R.Lebovitz, W.H.Reid and P.O.Vandervoort (Eds.) (University of Chicago Press,
Chicago 1978).

\bigskip
\noindent
27. P.Haensel, J.L.Zdunik and R.Schaeffer, {\it Astron.Astrophys.} {\bf 160},
251 (1986).

\bigskip
\noindent
28. R.E.Lingenfelter and J.C.Higdon, {\it Nature} {\bf 356}, 132 (1992) ;
J.C.Higdon and R.E.Lingenfelter, in Ref.2, 586.

\bigskip
\noindent
29. I.A.Smith and D.Q.Lamb, {\it Astrophys.J.} {\bf 410}, L23 (1993).

\bigskip
\noindent
30. H.Li, R.Duncan and C.Thompson, in Ref.2, 600.

\bigskip
\noindent
31. J.Hakkila et al., in Ref.2, 59.

\bigskip
\noindent
32. I.A.Smith, in Ref.2, 610.

\bigskip
\noindent
33. D.Eichler, in Ref.2, 54.

\bigskip
\noindent
34. R.N.Manchester and J.H.Taylor, {\it Pulsars} (Freeman, San Francisco 1977).

\bigskip
\noindent
35. T.L.Cline et al., {\it Astrophys.J.} {\bf 255}, L45 (1982).

\bigskip
\noindent
36. R.Ramaty et al., {\it Nature} {\bf 287}, 122 (1980).

\bigskip
\noindent
37. D.A.Frail and S.R.Kulkarni, in Ref.2, 486.

\bigskip
\noindent
38. S.R.Kulkarni and D.A.Frail, {\it Nature} {\bf 365}, 33 (1993).

\bigskip
\noindent
39. T.Murkami et al., in Ref.2, 489.

\bigskip
\noindent
40. J.P.Norris et al., in Ref. 13, 479.

\bigskip
\noindent
41. R.C.Duncan and C.Thompson, in Ref.2, 625.

\bigskip
\noindent
42. B.Schutz, in Ref. 10, 191.

\bigskip
\noindent
43. D.Nicholson and B.Schutz, in Ref. 10, 388.

\end